\documentclass[aps,prl,twocolumn]{revtex4}
 \usepackage{amsmath}
\usepackage{amsfonts}
\usepackage{amssymb}
\usepackage{dcolumn}
\usepackage{bm}
\usepackage{graphicx}
\usepackage{epstopdf}
\usepackage{color} 
\usepackage{ulem}

\def \C {$^{13}$C }
\def \tt {$ \left (T_{1}T \right) ^{-1}$ } 
\def \ttE {$ \left (T_{1}T \right) ^{-1}$}     

 \def\KEt {\mbox{$\kappa$-(BEDT-TTF)$_2$Cu(NCS)$_2$}}
    \def\KEtS {\mbox{$\kappa$-(BEDT-TTF)$_2$Cu(NCS)$_2$ }}
  \def\KE {\mbox{$\kappa$-(ET)$_2$X}}
   \def\KES {\mbox{$\kappa$-(ET)$_2$X }}

\begin{document}

\title{Evidence of   Andreev bound states  as a hallmark of   the FFLO phase in \KEt   }

\author{ H. Mayaffre$^{1}$, S. Kr{\"a}mer$^{1}$, M. Horvati{\'c}$^{1}$, C. Berthier$^{1}$, K. Miyagawa$^{2}$ , K. Kanoda$^{2}$,    
  and  V. F. Mitrovi{\'c}$^{3,*}$}
  
\address{
\mbox{$^{1}$Laboratoire National des Champs Magn{\' e}tiques Intenses, LNCMI - CNRS  (UPR 3228),  } UJF, UPS and INSA, BP 166, 38042 Grenoble Cedex 9, France\\
\mbox{$^{2}$Department of Applied Physics, University of Tokyo, Bunkyo-ku, Tokyo 113-8656, Japan}\\
\mbox{$^{3}$Department of Physics, Brown University, Providence, RI 02912, U.S.A.} \\
 $^{*}$e-mail: vemi@brown.edu}

\pacs{ 74.70.Tx, 76.60.Cq, 74.25.Dw, 71.27.+a }
\maketitle


{\bf 
Superconductivity is a quantum phenomena arising, in its simplest form, from pairing of  fermions with opposite spin into a state with zero net momentum.  Whether superconductivity can occur in fermionic systems with unequal number of two species distinguished by 
 spin, atomic hyperfine states, flavor, presents an important open question in condensed matter, cold atoms, and quantum chromodynamics, physics. In the former case the imbalance between spin-up and spin-down electrons forming the Cooper pairs is indyced by the magnetic field. Nearly fifty years ago Fulde, Ferrell, Larkin and Ovchinnikov (FFLO) proposed that such imbalanced system can lead to exotic superconductivity in which pairs acquire finite momentum \cite{Fulde}. The finite pair momentum leads to spatially inhomogeneous state consisting of 
  of a periodic alternation of ``normal'' and ``superconducting'' regions.  
 Here, we report nuclear magnetic resonance (NMR) measurements providing  microscopic evidence for the existence of this new superconducting state through the observation of   spin-polarized quasiparticles forming so-called Andreev bound states. 
   }

\begin{figure}[t]
\begin{minipage}{0.98\hsize}
\centerline{\includegraphics[scale=0.45]{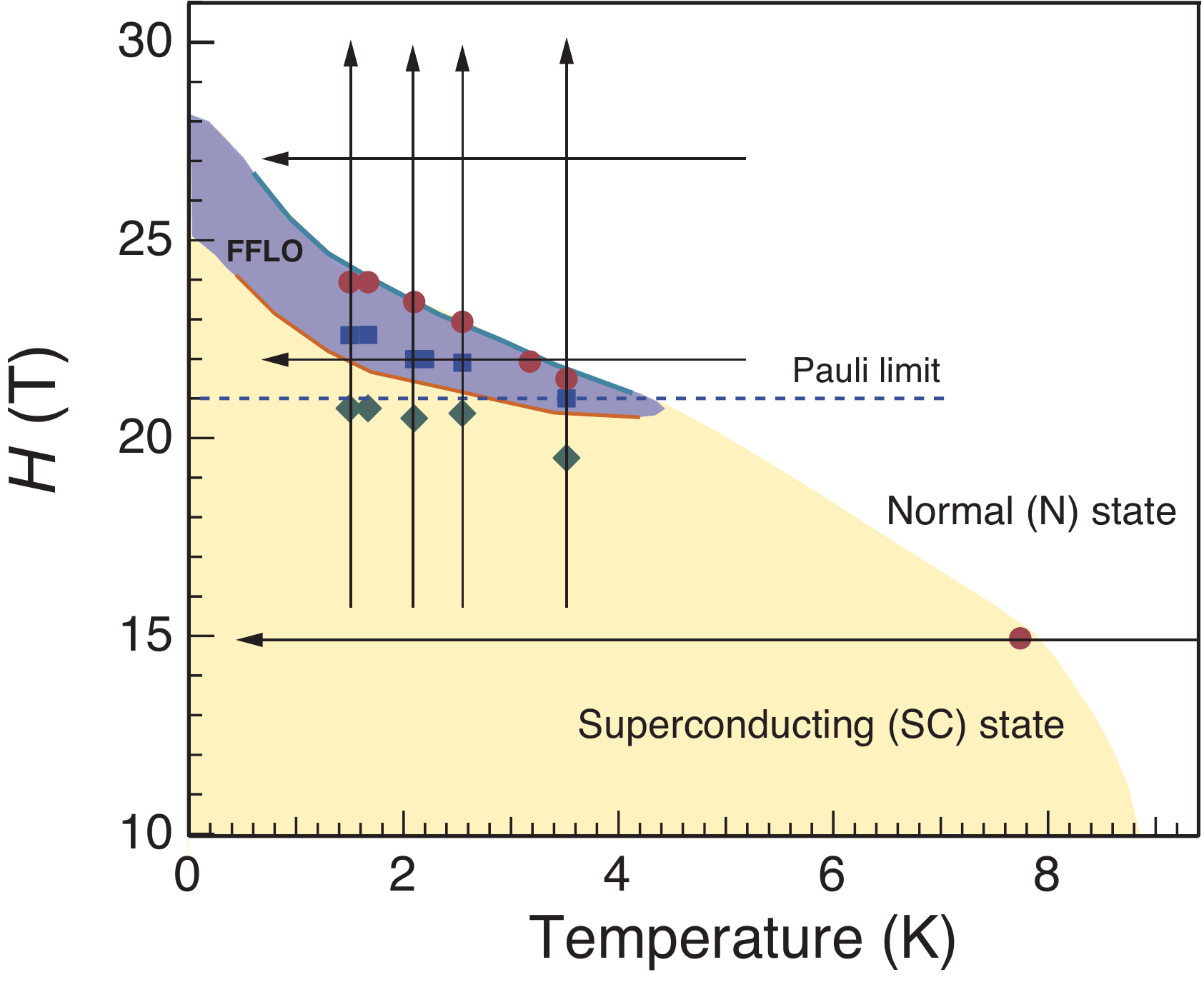}} 
\begin{minipage}{1\hsize}
 \vspace*{-0.2cm}
\caption[]{\label{Fig1}{\bf Phase diagram of $\kappa$-(BEDT-TTF)$_2$Cu(NCS)$_2$.} 
 }
 \vspace*{-0.7cm}
\end{minipage}
\end{minipage}
\end{figure}

The FFLO phase is expected to occur in the vicinity of 
 the upper critical field $(H_{c2})$ when Pauli pair breaking dominates over orbital (vortex)
effects \cite{Fulde,  Maki66, Gruenberg66}. Pauli pair breaking prevails in fields that exceed so-called Pauli limit $(H_{\rm p})$ 
 for which  the Zeeman energy is strong enough to break the Cooper pair by flipping one spin of the singlet and so destroy superconductivity. Intense efforts have been invested to search for indisputable evidence for the existence of the FFLO states.
Examples include theoretical proposal for detecting modulated superfluid phase in optical
lattices \cite{Loh10}; study of the tunneling density of states of superconducting films in high magnetic
fields \cite{Loh11};  mapping of the phase diagram of CeCoIn$_{5}$ \cite{Bianchi02, Koutroulakis10}, and  studies of layered organic
 superconductors \cite{Singleton00, Lortz07, Bergk11,Uji06}.   However clear microscopic evidence is still missing. 
 Besides CeCoIn$_{5}$, where putative FFLO state coexists with long range magnetism,   organic compound,  \KEtS (hereafter referred as  \KE) exhibits  the clearest thermodynamic evidence for the existence of
 a narrow intermediate SC  phase \cite{Lortz07} for the in-plane orientation of magnetic field that eliminates vortex effect.
Since this SC phase is stabilized in magnetic fields that exceed
Pauli limit, $H_{\rm p} \approx 20.7 \, {\rm T}$ \cite{Agosta12}, as
illustrated  in \mbox{Fig. \ref{Fig1}}, it has been identified as
an FFLO phase.   Recent measurements of nuclear magnetic resonance (NMR) spectra gave evidence that the phase transition within the SC state is Zeeman-driven \cite{Wright11}, but failed to provide clear hallmark of the FFLO state. Our main  discovery  is that NMR
relaxation rate becomes  significantly enhanced,  as compared to its normal state value, in the SC state for fields exceeding $H_{\rm p}$.  We deduce that the enhancement stems from the Andreev bound states  (ABS) of polarized quasiparticles 
spatially localized in the nodes of the order parameter in an FFLO
state and so identify microscopic nature of this high field phase. 

%
 %
%

\begin{figure}[t]
\begin{minipage}{1\hsize}
\centerline{\includegraphics[scale=0.75]{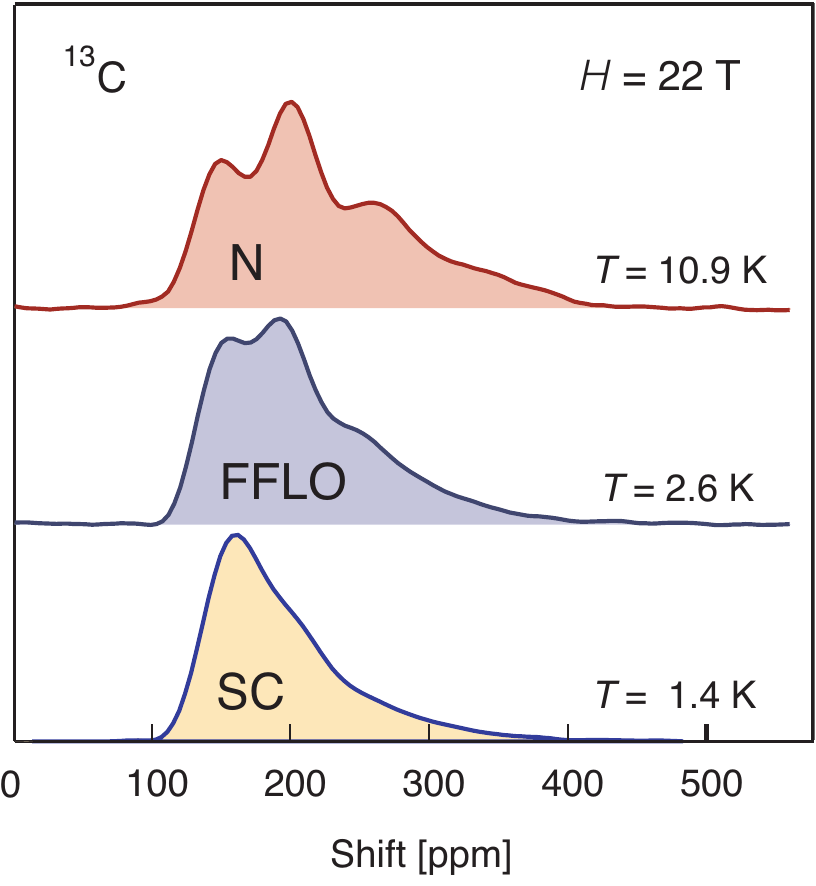}} 
\begin{minipage}{.98\hsize}
 \vspace*{-0.2cm}
\caption[]{\label{Fig1a}   {\bf  High field spectra  of \KE.}   }
 \vspace*{-0.4cm}
\end{minipage}
\end{minipage}
\end{figure}

We first examine the NMR spectral shapes displayed in   \mbox{Fig. \ref{Fig1a}}, in different regimes at 22 T, to   demonstrate sensitivity of our measurements  to different superconducting phases in this compound. 
These $^{13}$C NMR spectra reflect the distribution of the
hyperfine fields and are thus sensitive probe of the electronic
spin polarization \cite{Wright11, Kawamoto95, Mayaffre95}. In the
normal state at \mbox{10.9 K} the spectrum is relatively broad and
displays multiple peaks corresponding to inequivalent $^{13}$C
spin-labelled sites \cite{Kawamoto95,Mayaffre94,desoto95}. 
At \mbox{1.4 K}, deep in the superconducting  state, the spectrum is significantly narrower   than in the
normal state. This  narrowing is due to the decrease of the average electronic spin
polarization   in the superconducting  state, which in turn reduces the
splitting between the lines corresponding to distinct carbon
sites. At \mbox{2.6 K},  in putative FFLO state, the spectrum is narrower than in the
normal state yet wider than deep in the superconducting  state. 
 This indicates that  the sample is indeed in the SC state at this
$T$ and that the   electronic density of states  (DOS) at the Fermi energy $(E_{F})$ is suppressed below $T_{c}$. 
Thus, whatever the nature of the high field superconducting state is, its DOS at $E_{F}$ (averaged over the sample volume)  is suppressed as compared to the normal state.

%

\begin{figure}[t]
\begin{minipage}{0.98\hsize}
\centerline{\includegraphics[scale=0.45]{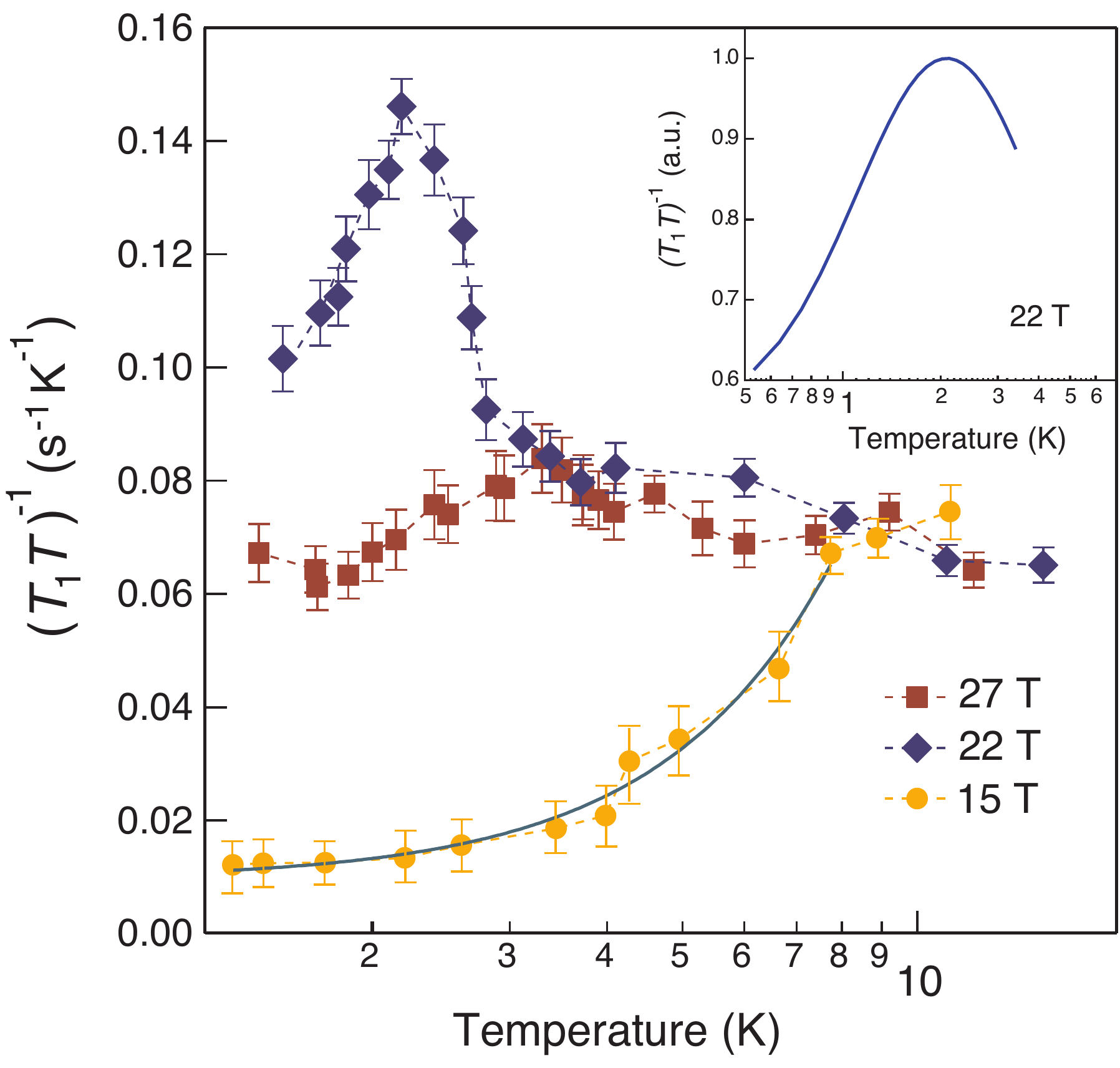}} 
\begin{minipage}{.98\hsize}
 \vspace*{-0.1cm}
\caption[]{\label{Fig2} 
 {\bf NMR relaxation rate in the normal and superconducting states.}
 }
 \vspace*{-0.3cm}
\end{minipage}
\end{minipage}
\end{figure}

Temperature   dependence  of   \tt    measured at various magnetic fields,
applied in the conducting planes,  is plotted in \mbox{Fig. \ref{Fig2}}. In the normal state \tt is constant,
 indicating that the  DOS 
in the vicinity of Fermi level is constant (Supplementary Information) \cite{AbragamBook}.  
At \mbox{27 T} \tt remains nearly constant within the error bars, and, consequently, implying that 
the sample remains in the normal state down to the lowest $T$ $(1.3 \, {\rm K})$ that we investigated,
 At \mbox{15 T}, the rate   decreases below $ T_{C} \approx 8 \, {\rm K}$  evidencing formation of the singlet SC state.
  At intermediate field of \mbox{22 T}, the sample remains
  in the normal state down to $T_{c} \approx 3\, {\rm K}$. 
 Below \mbox{$T_{c}$}, \tt sharply increases  to nearly twice the normal state value,
 reaching a maximum in the vicinity of  \mbox{2 K}.

%
%
 %
%

\begin{figure}[t]
\begin{minipage}{0.98\hsize}
\centerline{\includegraphics[scale=0.40]{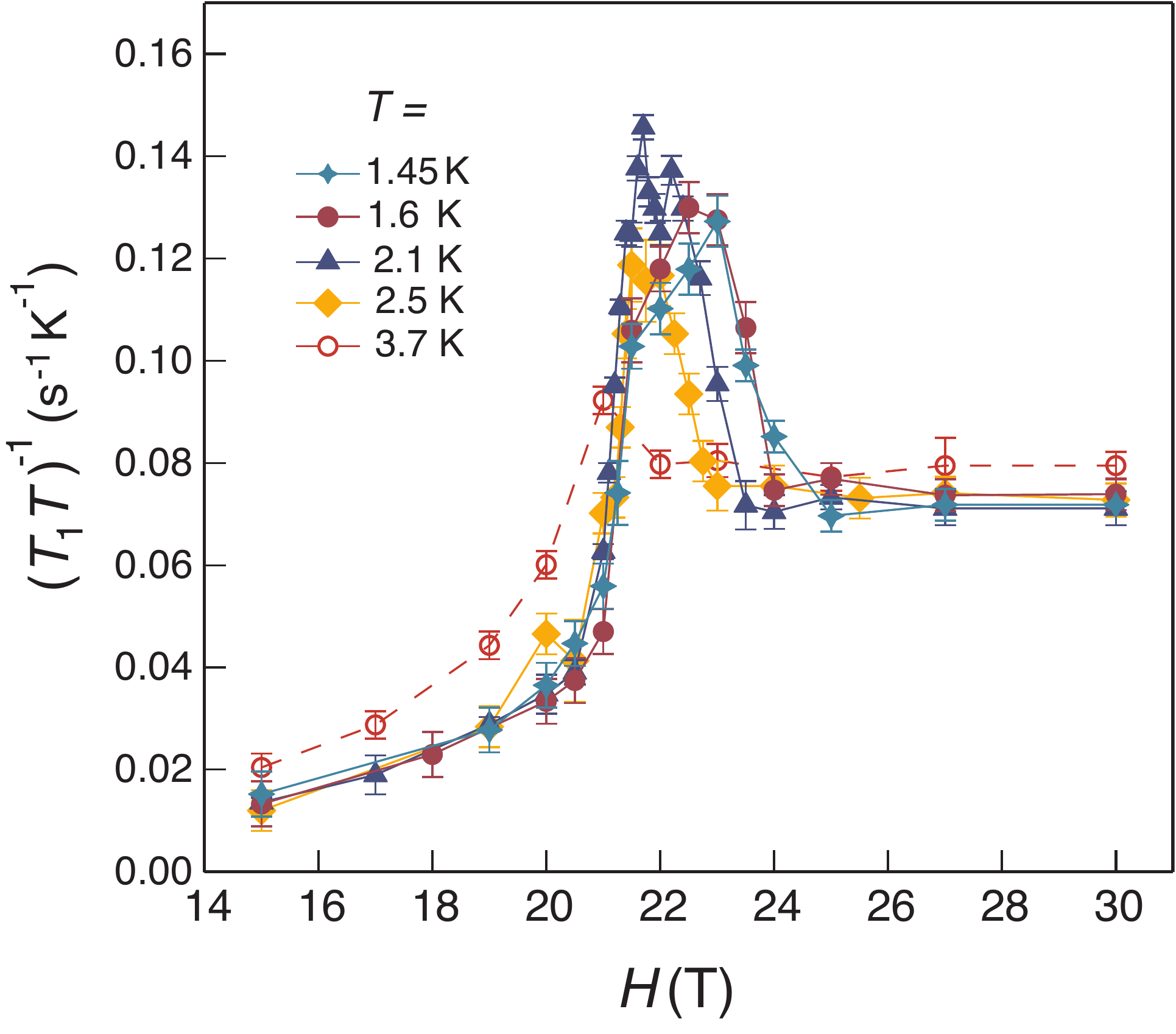}} 
\begin{minipage}{.98\hsize}
 \vspace*{-0.3cm}
\caption[]{\label{Fig3} 
{\bf Enhancement of the NMR relaxation rate in the FFLO state.} }
 \vspace*{-0.6cm}
\end{minipage}
\end{minipage}
\end{figure}

The field dependence of \tt at various temperatures as plotted in \mbox{Fig. \ref{Fig3}}.
In the normal state above $25 \; {\rm T}$,   \tt is applied field $(H)$
and $T$  independent.   Below $21 \; {\rm T}$, \tt is
suppressed below its normal state value due to the
formation of singlet Cooper pairs  in the SC state. The
enhancement of \tt is observed for the intermediate  field values from
$21$ to $24 \; {\rm T}$.  The observed enhancement of \tt is stunning because it  only appears in the SC state in fields exceeding $H_{\rm p}$ and where spectral measurements indicate suppression of the DOS at the $E_{F}$.

An important question is whether the observed enhancement of \tt in a SC state can be explained by more `standard'  mechanisms  amplifying the NMR relaxation in a SC state, e.g.  due to vortices. 
In the following we show that   such mechanisms can be readily discarded (see also Supplementary Information). Vortices can provide two
relaxation channels: one due to quasi-particles and the other to
field fluctuations induced by vortex  motion. 
Our measurements have been carried out for the applied field being precisely parallel to the conducting planes, as 
 identified
by the minimum in $T_1^{-1}$ (Supplementary Information). This minimum implies that only
Josephson-like vortices exist; these `coreless' vortices
\cite{clem90} contain no quasi-particles and thus
do not contribute to $T_1^{-1}$ \cite{Mayaffre95,desoto93} (Methods).  As for the vortex motion, the Josephson  vortices  
are formed in the insulating layers and strictly speaking magnetic
flux lines do not penetrate the SC layers \cite{Mansky93}. Thus,
even though these vortices are only weakly pinned \cite{Mansky94}, their
motion should not significantly contribute to the \C $T_1^{-1}$ rate,  
dominated by the hyperfine coupling to the SC-plane's electronic
degrees of freedom \cite{Mansky93} (Supplementary Information).

 %
%

 In the absence of magnetic correlation, the NMR relaxation rate is given by, 
 \begin{equation}
 \label{eqT1}
{1 \over T_{1}T} \propto {1  \over T}   \int d\varepsilon  \, N_{\uparrow} (\varepsilon) N_{\downarrow}  
 (\varepsilon ) f(\varepsilon ) [1 - f(\varepsilon  )],
\end{equation}
 where $N_{\downarrow}$ and $N_{\uparrow}$ denote the densities  of up- and down-spin states   and  $f(\varepsilon)$ is the Fermi occupation function. That is, \tt  is proportional to the square of the DOS  averaged
over energy $\varepsilon$ in a range of the order of $k_B T$ around   $E_F$. 
This implies that in a superconducting state with the 
 $d$-wave gap symmetry in the excitation spectrum, at low $T$
 only the DOS in the regions around the nodes of the gap contribute  to  
the relaxation rate.   For energies less than half of the gap magnitude, 
the DOS near  the nodes depends linearly on quasiparticle excitation energy leading to well known   $T^{2}$ dependence of the \ttE, illustrated in \mbox{Fig. \ref{Fig2}}. This quadratic decrease of \tt with decreasing $T$ in the $d$-wave SC state  is a direct consequence of the lack of low energy features in the DOS.  More complex   temperature dependence of the relaxation rate, as that observed in fields above \mbox{$H_{\rm p} \approx 21$ T}, can be generated by a   complex  peak-like structure in the quasiparticle  DOS at low energies around the $E_{F}$. 
The question remains as to what gives rise to such DOS features, {\it i.e.} bound states. The bound states  can only form in regions where SC order parameter is suppressed.  As presence of vortex cores  in our experiment was excluded, 
the ABS formed near the zeros of the FFLO order parameter in a $d$-wave superconductor   \cite{Vorontsov05,   {MakiPeak}, Wang06, Yanase09, Cui12} provide a natural explanation.   In fact, a sharp peak-like structure around $E_{F}$    is predicted in the FFLO state  in the vicinity of the transition from the SC to FFLO state, where nodes in the order parameter form the domain walls \cite{Vorontsov05}.  Qualitative  $T$ dependence of \tt arising from such peak-like DOS structure is shown in the inset to \mbox{Fig. \ref{Fig2}}. Evidently, it  is    in sharp contrast with previously discussed standard behavior for a 
 $d$-wave  SC state at lower field. 

%
 %
\begin{figure}[h]
\begin{minipage}{0.98\hsize}
\centerline{\includegraphics[scale=0.45]{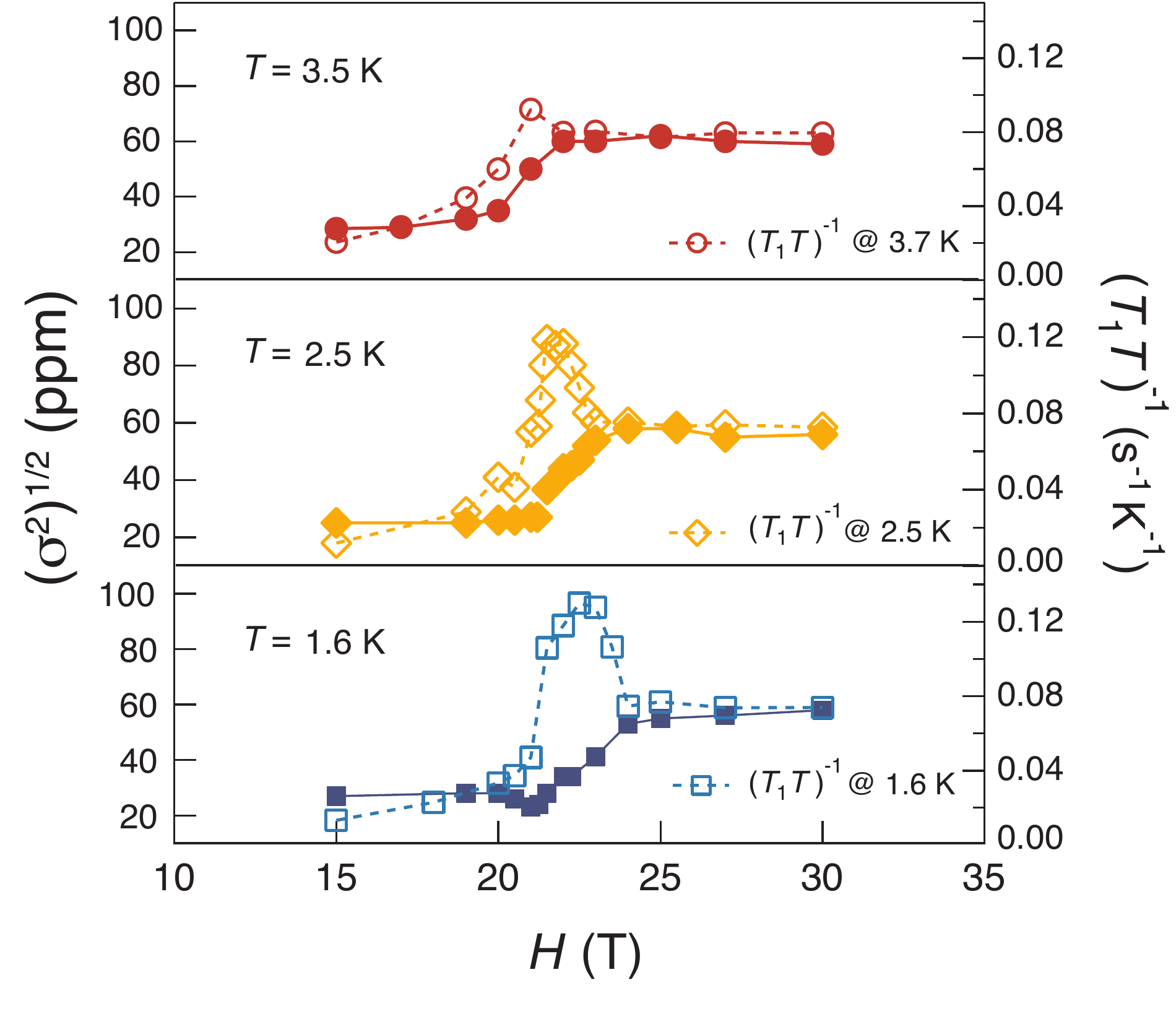}} 
\begin{minipage}{.98\hsize}
 \vspace*{-0.3cm}
\caption[]{\label{Fig4} 
{\bf Field dependence of the electronic spin polarization and NMR relaxation rate at low temperatures.}}
 \vspace*{-0.4cm}
\end{minipage}
\end{minipage}
\end{figure}
%

   To emphasize our main finding that the observed enhancement of \tt in the high field superconducting state indicates the presence of the ABS formed near the zeroes of the FFLO order parameter, in \mbox{Fig. \ref{Fig4}}
 we plot the square root of the second moment of the lineshape, as a quantitative measure of the width of  inhomogeneous spectra, together with the \tt data. 
  It is evident that the onset of the \tt enhancement is concurrent with the spectral line narrowing generated by the
  decreasing electronic spin polarization (and thus DOS at $E_{F}$) in the SC state.  
That is, both DOS at $E_{F}$ and the average spin polarization are lower than in  the normal
 state. Thus, the \tt enhancement over the normal state value  is  to be assigned exclusively to  {\it quasi-particle bound states located away from $E_{F}$}. 
 Such  bound states shifted away from the $E_{F}$, forming in applied fields exceeding $H_{\rm p}$, are the hallmark of an 
 FFLO state  \cite{Vorontsov05}.  Since these states are localized in the real space in the nodal region, they occupy a small fraction of the order of several percent of the total sample volume;  even if they were to produce finite DOS at $E_{F}$,   they will not contribute to any significant broadening or shift of the NMR spectra which reflect  the average over the entire sample. 
 However, such localized DOS can affect the global NMR rate due to nuclear spin  diffusion,  ``transferring'' the effect from nuclei spatially localized in the nodes to those far outside as well. The fact that the relaxation profile is homogeneous across the NMR spectra and follows  pure exponential time recovery confirms that fast spin diffusion is indeed effective.  

The increasing \tt with decreasing $T$  is a direct consequence of the appearance of the sharp (compared to $T$) peak-like DOS  (bound state) structure located away from $E_{F}$. The energy position and the width of these bound states  is set by  the applied field, as described in detail in the next paragraph. 
At a given $H$, the only effect of $T$ on \tt is to vary   the range $\varepsilon \simeq k_B T$    around $E_{F}$ over which the square of the DOS is averaged. 
Thus, on lowering $T$ in the FFLO state at fixed $H$,   \tt first increases as the bond states are   created  just below $T_c$. 
For temperatures below which    the bound state DOS energy 
exceeds  $\varepsilon$, \tt will decrease. This is in qualitative agreement with the observed $T$  dependence of \tt  at \mbox{22 T}, in the FFLO state, plotted in \mbox{Fig. \ref{Fig2}}.  

Quantitative explanation and comparison of the field dependence is more difficult as the effect of $H$ is twofold. 
In addition to shifting the bound states away from the $E_{F}$,  the applied field  
controls the sharpness of these states. That is, as the field increases and  more nodal planes are introduced, bound states broaden in energy due to hybridization of the energy levels corresponding to the adjacent planes. As 
 a result the sharp bound state for a single domain wall (formed near the transition from the  SC state) broadens 
 and thus fills the low energy region below the maximum gap. 
 At a given $T$, it is the interplay of the broadening and the peak energy of the  bound states, both controlled by the $H$, that determines   the 
 $H$ value at which the peak in \tt is observed (Methods).

 In addition to the contribution to   \tt related to the enhancement of DOS by the Andreev bound states,
   we cannot {\it a priori} discard contribution from  
 {\it dynamical} effects associated with these same bound states.
 The corresponding polarized  quasiparticles  
   could tunnel between different nodal planes of the order parameter and produce
  fluctuating fields responsible for nuclear relaxation.   Another   possible  source of fluctuating fields is motion of the nodal planes (Supplementary Information)  
  in which bound states of polarized quasiparticles are localized
 \cite{Loh10}.  
 Precise calculations of these  dynamical  contributions   to \tt are complex and
  have not yet been performed.      \\

{\bf   METHODS}\\
   
 {\bf Sample \& NMR methods.} Experiment were performed on high quality   \KES single crystals,
grown by electrolytic method \cite{Urayama88}. The most suitable nucleus for
our NMR study is $^{13}$C. As  it has a low natural abundance,
 we used samples  selectively enriched with  $^{13}$C on the site that is the most sensitive to electronic degrees
  of freedom. Such   sites  are  located on the central C-C pair that bears the largest spin density in the molecule
  \cite{Kawamoto95}. 
  We used samples with \mbox {100 \%}  $^{13}$C enriched pairs, giving rise to eight  NMR inequivalent sites.  
   The sample, placed inside the radio frequency   NMR coil,  was oriented by an accurate mechanical  goniometer. 
   As the field must be applied strictly parallel to the conducting  planes with a precision  better than $1.4^{\circ}$
   for the possible FFLO phase to be stabilized \cite{Bergk11}, 
  sharp minimum in the $T_{1}^{{-1}}$  was used as   sensitive {\it in-situ} signature  of the precise alignment of
  the field within conduction planes \cite{desoto93} (see also Supplementary Information).   That is, when the field is exactly aligned  along the  conducting planes
  only Josephson-type vortices can form. In the core of such vortices   quasiparticles are depleted leading to a significant suppression of $T_{1}^{{-1}}$  
   \cite{Mayaffre95,desoto93}.

  The measurements were done at the LNCMI in Grenoble,  using  
superconducting magnet \mbox{(for $H  = 15\, {\rm T}$)}    and
resistive magnet at higher fields. The temperature control was provided by $^4$He
variable temperature  insert. The NMR data were recorded using a
state-of-the-art laboratory-made NMR spectrometer. $T_1^{-1}$ was
measured by the saturation-recovery method: following the  saturation of nuclear magnetization  obtained
 by applying  a train of  $\pi/2$ pulses
equally spaced by a time $t  \geq T_2$, the signal was detected after a variable delay time
using a standard  spin echo sequence $(\pi/2-\tau-\pi)$. \\

 {\bf $T_{c}$ determination.}  $T_{c}$ was identified by   examining the NMR spectral
  shapes and shifts, and  the tuning resonance of the NMR tank circuit. \\

  {\bf Supplementary Information}  and any associated references are available in the online version of the paper at www.nature.com/nature.
   
  \vspace{-0.2cm}
\bibliographystyle{unsrt}

\vspace{0.9cm}

{\bf Figure Legends:} \\

{\bf 1.} {\bf Phase diagram of $\kappa$-(BEDT-TTF)$_2$Cu(NCS)$_2$.} 
Curves and  color shaded areas sketch
  the $H$-$T$ phase
diagram based on magnetic torque measurements
\cite{Bergk11} for $H$ parallel to the conducting planes.
Circles denote $T_{c}$, transition temperature from the normal to
the SC state, determined from our NMR data as explained in the
text, while squares  mark the peak in the NMR rate. Diamonds
denote the onset $H$ above which \tt exceeds the value
extrapolated from the low field SC state below Pauli limiting field of
$\approx 20.7\, {\rm T}$ \cite{Agosta12}. Arrows indicate
the field  and temperature ranges in which NMR relaxation rate   were taken. \\

{\bf 2.} {\bf  High field spectra  of \KE.}  \C NMR spectra   at  22 T field applied parallel to the conducting planes in the 
superconducting   $(T = 1.4 \, {\rm K})$, possible FFLO $(T = 2.6
\, {\rm K})$, and normal state $(T = 10.9 \, {\rm K})$.  Multiple peaks evident in the normal state spectrum arise from eight distinct crystallographic sites of $^{13}$C (see Methods section). \\

{\bf 3.} {\bf NMR relaxation rate in the normal and superconducting states.}
 Temperature dependence of \C  
NMR \tt   at  $H = $ 15 T, 22 T, and
\mbox{27 T}  $H$ applied in the conducting planes.
Solid line denotes the quadratic $T$ dependence characteristic for superconductors
  with  a gap having a line of nodes, such as  for a \mbox{$d$-wave} symmetry.
The dashed lines are guide to the eye.  The parts of the phase
diagram explored are shown by horizontal arrows in \mbox{Fig.
\ref{Fig1}}. Inset:   Calculated temperature dependence of the \tt  arising from the Andreev bound states formed at the nodes of the order parameter. The result is only qualitatively correct as states are phenomenologically modeled by gaussians 
 shifted away from the Fermi level by the amount proportional to the applied field \cite{Vorontsov05} and the $T$ dependence of the gap is neglected. \\

{\bf 4.} {\bf Enhancement of the NMR relaxation rate in the FFLO state.} NMR relaxation rate of
\C divided by $T$  at various temperatures as a function of
magnetic field applied in the conducting planes. The parts of the
phase diagram explored are shown by vertical arrows in \mbox{Fig. \ref{Fig1}}. \\

{\bf 5.} {\bf Field dependence of the electronic spin polarization and NMR relaxation rate at low temperatures.}
Square root of the second moment, measuring electronic spin polarization, of the \C NMR spectra (filled symbols)  as a function of magnetic field applied in the conducting planes at different $T$. Typical error bars are of the order of a few percent and not shown for clarity.  \tt (open symbols) field dependence at the corresponding temperatures is shown
for comparison. Lines are guide to the eye. \\

{\bf Acknowledgements}
We would like to thank   A. Vorontsov, Y. Yanase, and M. Sigrist for   illuminating discussions, and I. Sheikin for providing raw data for the  phase diagram.  This research is supported by the funds from   the French ANR grant 06-BLAN-0111,  the EuroMagNET II network under EU Contract No. 228043,   the
visiting faculty program of Universit{\'e} Joseph Fourier (V.F.M.), and ADVANCE HRD-0548311 (V.F.M.).\\

{\bf Author Contributions}
K.M. and K.K. prepared the samples. H.M.,   S.K., and V.F.M.  performed the experiments. S.K. and M.H. developed and operated the high-field NMR facility. H.M. created software for spectrometers. H.M. and V.F.M. analyzed the data. C.B. provided conceptual advice  and contributed to the planning of the project. H.M., C.B, M.H, and V.F.M. developed data interpretation. V.F.M. wrote the paper and supervised the project. All authors discussed the results and commented on and edited the manuscript. \\

{\bf Author Information}
 Reprints and permissions information is available at www.nature.com/reprints. The authors declare no competing financial interests. Readers are welcome to comment on the online version of this article at www.nature.com/nature. Correspondence and requests for materials should be addressed to V.F.M. (vemi@brown.edu). \\

\end{document}